 \documentclass[twocolumn,showpacs,preprintnumbers,amsmath,amssymb]{revtex4}
\usepackage{dcolumn}% Align table columns on decimal point

\begin{document}

\title
      {Calculation of $P,T$-odd effects in $^{205}$TlF including electron
       correlation}
\author{A.\ N.\ Petrov}
\email{anpetrov@pnpi.spb.ru}
\homepage{http://www.qchem.pnpi.spb.ru}
\affiliation{Petersburg Nuclear Physics Institute,
              Gatchina, Petersburg district 188300, Russia}
\author{N.\ S.\ Mosyagin}
\affiliation{Petersburg Nuclear Physics Institute,
              Gatchina, Petersburg district 188300, Russia}
\author{T.\ A.\ Isaev}
\affiliation{Petersburg Nuclear Physics Institute,
              Gatchina, Petersburg district 188300, Russia}
\author{A.\ V.\ Titov}
\affiliation{Petersburg Nuclear Physics Institute,
              Gatchina, Petersburg district 188300, Russia}
\author{V.\ F.\ Ezhov}
\affiliation{Petersburg Nuclear Physics Institute,
              Gatchina, Petersburg district 188300, Russia}
\author{E.\ Eliav}
\affiliation{School of Chemistry, Tel Aviv University,
             Tel Aviv 69978, Israel}
\author{U.\ Kaldor}
\affiliation{School of Chemistry, Tel Aviv University,
             Tel Aviv 69978, Israel}

\date{\today}

\begin{abstract}
 A method and codes for two-step correlation calculation of heavy-atom
 molecules have been developed, employing the generalized relativistic
 effective core potential (GRECP) and relativistic coupled cluster (RCC)
 methods at the first step, followed by nonvariational one-center restoration
 of proper four-component spinors in the heavy cores.  Electron correlation is
 included for the first time in an {\it ab initio} calculation of the
 interaction of the permanent $P,T$-odd proton electric dipole moment with the
 internal electromagnetic field in a molecule.  Inclusion of electron
 correlation by GRECP/RCC has a major effect on the $P,T$-odd parameters of
 $^{205}$TlF, decreasing $M$ by 17\%  and $X$ by 22\%.

\end{abstract}

\pacs{31.90.+s, 35.10.Wb, 31.20.-d}
\maketitle

\paragraph*{Introduction.}

 The measurement of permanent electric dipole moments (EDM) of elementary
 particles is highly important for the theory of $P,T$-odd interactions.
 Experiments performed so far have given only upper bounds for the EDMs.  The
 extraction of EDMs from measurements on molecules containing heavy atoms
 requires knowledge of nuclear and electronic properties of the molecule.
 High quality calculations of the relevant electronic properties are therefore
 essential for accurate determination of the EDMs
 \cite{ObzorPT,ObzorPT1}.

 Here we consider the interaction of the proton EDM with the internal
 electromagnetic field of the $^{205}$TlF molecule.  This molecule is one of
 the best candidates for proton EDM measurements.
 Following Hinds and Sandars \cite{Hinds}, the effective interaction with the
 proton EDM in TlF is written in the form

\begin{equation}
     H_{\rm eff}=(d^V+d^M) \vec{\sigma} _N \cdot \vec{\lambda}\ ,
 \label{interaction}
\end{equation}
 where $\vec{\sigma}_N$ is the Tl nuclear spin operator, $\vec{\lambda}$ is
 the unit vector along the internuclear axis $z$ from Tl to F, $d^V$ and $d^M$
 are constants corresponding to the volume and magnetic effects according to
 Schiff's theory \cite{Schiff}.  Hinds and Sandars showed \cite{Hinds} that the
 volume effect in a coordinate system centered on the Tl nucleus is given by

\begin{equation}
   d^V=-d_pXR\ ,
 \label{dv}
\end{equation}
 where $d_p$ is the proton EDM, $R$ is a factor determined by the nuclear
 structure of $^{205}$Tl,

\begin{equation}
   X=\frac{2\pi}{3} \left[
     \frac{\partial}{\partial z}\rho_{\psi}(\vec{r})
      \right] _{x,y,z=0}\ ,
  \label{X}
\end{equation}
 $\rho_{\psi}(\vec{r})$ is the electronic density calculated from the wave
 function $\psi$.  Keeping only the dominant diagonal terms of the
 two-electron operator for the magnetic effect (see \cite{ObzorPT1}) they have

\begin{equation}
   d^M = 2 \sqrt{2}d_p
   \left(
   \frac{\mu}{Z}+ \frac{1}{2mc}
   \right)M\ ,
 \label{dm}
\end{equation}
 where $\mu$, $m$ and $Z$ are the magnetic moment, mass and charge of the Tl
%
% nucleus correspondingly,
 nucleus,
 $c$ is the velocity of light,

\begin{equation}
   M = \frac{1}{\sqrt{2}}\langle\psi |\sum_i\left(\frac{\vec{\alpha}_i \times
                         \vec{\bf l}_i}{r_i^3}\right)_{z}|\psi \rangle\ ,
 \label{M}
\end{equation}
 $\vec{\bf l_i}$ is the orbital momentum operator of electron $i$, and
 $\vec{\alpha}_i$ are its Dirac matrices.
 $H_{\rm eff}$ leads to different hyperfine splitting of TlF in parallel and
 antiparallel electric and magnetic fields. The level shift $h\nu =
 4(d^V+d^M)\langle \vec{\sigma} _N \cdot \vec{\lambda} \rangle$ is measured
 experimentally (for the latest data see \cite{experimentTlF}; another
 experiment is now in preparation at the Petersburg Nuclear Physics
 Institute).

 The parameters $X$ of Eq.\ (\ref{X}) and $M$ of Eq.\ (\ref{M}) are determined
 by the electronic structure of the molecule. They were calculated recently
 for the $X0^+$ ground state of TlF by Parpia \cite{Parpia} and by Quiney et
 al.\ \cite{Quiney} using the Dirac-Hartree-Fock (DHF) method with large
 Gaussian basis sets (see Table \ref{result}).  No calculation which includes
 correlation effects is available. The main goal of the present work is to
 calculate the $X$ and $M$ parameters for the molecule with correlation
 included to high order.

\paragraph*{Methods.}

 The generalized relativistic effective core potential (GRECP) method
 \cite{GRECP} is applied to the TlF molecule.  A two-component electronic
 (pseudo)wave function is first obtained with the 21-electron GRECP
 \cite{Hg-Bi-Tup,Hg-Rn-Mos} for Tl, providing proper electronic density in the
 valence and outer core regions, followed by restoration of the proper shape
 of the four-component molecular spinors in the inner core region of Tl.
 Details of the method may be found
 elsewhere~\cite{GRECP,restoration,TitovYbF,MosyaginYbF}.

 The correlation spin-orbital basis set used consisted of 26$s$, 25$p$, 18$d$,
 12$f$, and 10$g$ Gaussian-type orbitals on Tl, contracted to $6s6p4d2f1g$.
 The basis was optimized in a series of atomic two-component GRECP
 calculations, with correlation included by the all-order relativistic coupled
 cluster (RCC) method \cite{RCC} with single and double excitations; the
 average energy of the two lowest states of the atom was minimized. The basis
 set generation procedure is described in Refs.\ \cite{basis,HgH-CC}.  The
 basis set was designed to describe correlation in the outer core $5s$ and $5p$
 shells of Tl, in addition to the $5d$ and valence shells.  While $5s$ and $5p$
 correlation may not be important for many of the chemical and physical
 properties of the atom, it is essential for describing properties coming from
 inner regions, including $P,T$-odd effects. The ($14s9p4d3f$)/\-[$4s3p2d1f$]
 basis set from the ANO-L library \cite{MOLCAS} is used for fluorine.

 A one-component self-consistent-field (SCF) calculation of the $(1\sigma
 \dots 7\sigma)^{14}\-(1\pi 2\pi3\pi)^{12}(1\delta )^4$ ground state of TlF is
 performed first, using the GRECP for Tl which simulates the interactions of
 the valence and outer core ($5s5p5d$) electrons with the inner core
 [Kr]$4d_{3/2}^44d_{5/2}^64f_{5/2}^64f_{7/2}^8$.  This is followed by
 two-component RCC calculations, with only single (RCC-S) or with single and
 double (RCC-SD) cluster amplitudes.  The RCC-S calculations with the
 spin-dependent GRECP operator take into account effects of spin-orbit
 interaction at the level of the one-configurational SCF-type method.  The
 RCC-SD calculations include, in addition, the most important electron
 correlation effects.

 The electron density obtained from the two-component GRECP/RCC (pseudo)wave
 function in the valence and outer core regions is very close to that of the
 corresponding all-electron four-component function.  The pseudospinors are
 smoothed in the inner core region \cite{GRECP}, so that the electronic
 density in this region is not correct.  The operators in equations (\ref{X})
 and (\ref{M}) are heavily concentrated near the nucleus, and are therefore
 strongly affected by the wave function in the inner region. The
 four-component molecular spinors must therefore be restored in the inner
 region of Tl.  All molecular spinors $\phi _{i}$ are restored as one-center
 expansions on the Tl nucleus, using the nonvariational restoration scheme
 (see \cite{GRECP,restoration,TitovYbF,MosyaginYbF} and references therein).

 The restoration is started by generating equivalent basis sets of atomic
 (one-center) four-component spinors
 $\left\{ \left( \begin{array}{c} f_{nlj}(r)\chi_{ljm} \\
  g_{nlj}(r)\chi_{l'jm} \\ \end{array} \right) \right\}$
 and two-component pseudospinors $\{\tilde f_{nlj}(r)\chi _{ljm} \}$ by atomic
 finite-difference all-electron DHF and two-component GRECP/SCF calculations of
 the same valence configurations of Tl and its ions.  Here $n$ is the principal
 quantum number, $j$ and $ m$ are the total electronic momentum and its
 projection on the internuclear axis, $l$ and $l'$ are the orbital momenta, and
 $l'{=}2j{-}l$.  The nucleus is modeled as a uniform charge distribution within
 a sphere with radius $r_{\rm nucl} = 7.1 {\rm fm} \equiv 1.34\times10^{-4}$
 a.u., whereas previous calculations employed a spherical Gaussian nuclear
 charge distribution \cite{Parpia,Quiney} (the root mean square radius in all
 calculations is 5.5 fm, in accord with the parametrization of Johnson and Soff
 \cite{Johnson}, and agrees with the experimental value 5.483 fm for the
 $^{205}$Tl nucleus \cite{Fricke}).  The all-electron four-component {\sc hfd}
 \cite{Tup} and two-component {\sc  grecp/hfj} \cite{Hg-Bi-Tup,Hg-Rn-Mos} codes
 were employed to generate the two equivalent [$15s12p12d8f$] numerical basis
 sets for restoration.  These sets, describing mainly the core region, are
 generated independently of the basis set for the molecular GRECP calculations
 discussed earlier.  The molecular pseudospinorbitals are then expanded in the
 basis set of one-center two-component atomic pseudospinors,

\begin{equation}
    \tilde {\phi} _{i}({\bf r}) \approx
    \sum_{l=0}^{L_{max}}\sum_{j=|l-1/2|}^{j=|l+1/2|} \sum_{n,m}
    c_{nljm}^{i}\tilde f_{nlj}(r)\chi _{ljm}\ .
 \label{expansion}
\end{equation}
 Note that for linear molecules only one value of $m$ survives in the sum for
 every ${\phi} _{i}$.  Finally, the two-component pseudospinors in the basis
 are replaced by the equivalent four-component spinors and the expansion
 coefficients from Eq.~(\ref{expansion}) are preserved
 \cite{restoration,TitovYbF,MosyaginYbF}:

\begin{equation} {\phi} _{i}({\bf r}) \approx
    \sum_{l=0}^{L_{\rm max}}\sum_{j=|l-1/2|}^{j=|l+1/2|} \sum_{n,m}
    c_{nljm}^{i}
    \left(
    \begin{array}{c}
    f_{nlj}(r)\chi _{ljm}\\
    g_{nlj}(r)\chi _{l'jm}
    \end{array}
    \right)\ .
  \label{restoration}
\end{equation}
 The molecular four-component spinors constructed this way are orthogonal to
 the inner core spinors of Tl, as the atomic basis functions used in
 Eq.~(\ref{restoration}) are generated with the inner core electrons treated
 as frozen.

 The quality of the approximation for the two-center molecular spinors and,
 consequently, of the calculated properties increases with the value of
 $L_{\rm max}$. A series of calculations of the $M$ parameter was performed
 using Eq.\ (\ref{restoration}) with basis functions going up to $p$, $d$ and
 $f$ harmonics. We found (see Table \ref{result}) that including only $s$ and
 $p$ functions in the expansion determines $M$ with 90\% accuracy.  Because
 the contribution of $f$ is only about 0.3\% and amplitudes of higher
 harmonics on the nucleus are suppressed by the leading term
 $\sim r^{(j{-}1/2)}$, the error due to the neglect of spherical harmonics
 beyond $f$ is estimated to be below 0.1\%.  Calculation of the $X$ parameter
 requires $s$ and $p$ harmonics (see Ref.\ \cite{Quiney}),
 although, strictly speaking, $d$ harmonics also give nonzero contributions.

 The restoration procedure implemented here gives a very good description of
 the wave function in the core region, which is important for accurate
 evaluation of the $X$ and $M$ parameters. This is done at a fraction of the
 cost necessary for all-electron four-component molecular calculations with
 Gaussian basis sets, where a large number of additional basis functions must
 be included for proper description of the inner core region and small
 components of spinors \cite{Quiney}. Here we calculate (restore) the
 four-component electronic wave function in the core region from the
 (pseudo)wave function obtained in the molecular GRECP calculation, which may
 be considered ``frozen'' in the valence region at the restoration stage.
 Basis functions describing a large number of chemically inert core electrons
 may thus be excluded from the molecular GRECP calculation.

 The $X$ and $M$ parameters were calculated by the finite field method (see,
 e.g., Refs.\ \cite{kaldor,Monkhorst}).  The operator corresponding to a
 desired property [see Eqs.~(\ref{X}) and (\ref{M})] is multiplied by a small
 parameter $\lambda$ and added to the Hamiltonian.  The derivative of the
 energy with respect to $\lambda$ gives the computed property.  This is
 strictly correct only at the limit of vanishing $\lambda$, but it is usually
 possible to find a range of $\lambda$ values where the energy is linear with
 $\lambda$ and energy changes are large enough to allow sufficient precision.
 The quadratic dependence on $\lambda$ is eliminated in the present
 calculations by averaging absolute energy changes obtained with $\lambda$ of
 opposite signs.

\paragraph*{Results and discussion.}

 Calculations were carried out at two internuclear separations, the
 equilibrium $R_e=2.0844$ \AA\, as in Ref.\ \cite{Parpia}, and 2.1 \AA, for
 comparison with Ref.\ \cite{Quiney}.  The results are collected in Table
 \ref{result}. The first point to notice is the difference between
 spin-averaged SCF and RCC-S values, which include spin-orbit interaction
 effects. These effects increase $X$ by 9\% and decrease $M$ by 21\%.  The
 RCC-S function may be written as a single determinant, and results may
 therefore be compared with DHF values, even though the RCC-S function is not
 variational. GRECP/RCC-S values of the $M$ parameter are indeed within 3\% and
 1\% of corresponding DHF values \cite{Parpia,Quiney} (Table \ref{result}).
 This agreement confirms the validity of the approximations made by us.  In
 particular, freezing the inner core shells is justified, as inner core
 relaxation effects have little influence on the properties calculated here, a
 conclusion already drawn by Quiney et al.\ \cite{Quiney}.

 Much larger differences occur for the $X$ parameter.  Here there are also
 large differences between the two DHF calculations, which cannot be explained
 by the small change in internuclear separation.  The value of $X$ may be
 expected to be less stable than $M$, because it is determined by the
 derivative of the electronic density at the Tl nucleus and involves large
 cancellations \cite{Quiney} between contributions of large and small
 components, each of which is about 20 times larger than their sum.  Thus, a
 strong dependence of $X$ on the basis used may be expected.  The DHF values
 collected in Table \ref{result} indeed show such dependence. Results obtained
 in Refs.\ \cite{Parpia} and \cite{Quiney} with comparable even-tempered basis
 sets, $(28s28p12d8f)$ and $(28s28p14d8f)$, are rather close, differing by 340
 a.u. Improving the Tl basis to $(34s34p16d9f)$ \cite{Quiney} increases $X$ by
 650 a.u.\ or 8\%.  Further improvement of the basis may be expected to yield
 even higher $X$ values.  The numerical basis functions obtained in atomic DHF
 calculations and used for the restoration are highly accurate near the
 nucleus, so that our RCC-S value for $X$, which is higher than that of Quiney
 et al.\ \cite{Quiney}, seems reasonable.  The different nuclear models used
 in the present and DHF \cite{Parpia,Quiney} calculations may also contribute
 to the disagreement in $X$, which is determined by the derivative of the
 electronic charge density at a single point, the Tl origin. $M$ is affected
 by $\psi$ in a broader region, and is therefore far less sensitive to the
 nuclear model.

 The main goal of this work is the evaluation of electron correlation effects
 on the $P,T$-odd parameters. These effects are calculated by the RCC-SD
 method at the molecular equilibrium separation $R_e$. A major correlation
 contribution is observed, decreasing $M$ by 17\% and $X$ by 22\%.

 Using the correlated values for $X$ and $M$ calculated here and
 $R=1.036\times10^{-9}$ a.u. from \cite{Coveney},
 one obtains from Eqs.\ (\ref{dv}) and (\ref{dm})

\begin{equation}
   d^V=-7.909\times10^{-6}d_p\ {\rm a.u.}
 \label{dvr}
\end{equation}
\begin{equation}
   d^M=1.622\times10^{-6}d_p\ {\rm a.u.}
 \label{dmr}
\end{equation}
 The effective electric field interacting with the EDM of the valence proton of
 $^{205}$Tl in the fully polarized TlF molecule is
 $E=|d^V+d^M|/d_p=6.287{\times}10^{-6}\ {\rm a.u.} = 32.33$ kV/cm;
 the revised proton EDM limit for the experiment of Ref.\ \cite{experimentTlF}
 is $d_p = (-1.7 \pm 2.8){\times}10^{-23}\ {\rm e{\cdot}cm}$.

 The hyperfine structure constants of Tl $6p_{1/2}^1$ and Tl$^{2+}$ $6s^1$,
 which (like $X$ and $M$) depend on operators concentrated near the Tl
 nucleus, were also calculated. The errors in the DF values are 10--15\%;
 RCC-SD results are within 1--4\% of experiment. The improvement in $X$ and
 $M$ upon inclusion of correlation is expected to be similar.

\paragraph*{Concluding remarks.}

 Note that the codes developed for GRECP/RCC calculation followed by
 nonvariational one-center restoration in heavy cores are equally
 applicable to calculation of other properties described by operators
 singular near nuclei (hyperfine structure, quantum electodynamic effects,
 etc.).
 Because the Fock-space RCC-SD approach \cite{RCC} is used, the two-step
 method is applicable to both closed-shell and open-shell systems, including
 excited states. In particular, calculations for the ground state of YbF
and for excited states of PbO are in progress now.
 Triple and higher cluster amplitudes in the valence region are important for
 chemical and spectroscopic properties, but not for the effects discussed
 here, as concluded from previous calculations for YbF \cite{TitovYbF}.
 These excitations are believed to be unimportant in the core region too.  We
 therefore suggest that further improvement in the correlation treatment will
 not seriously affect our $M$ and $X$ values.
%<commented for brevity>:
% One more remark is concerning the different HFD versions used by Parpia
% (unrestricted HFD) and by Quiney et al.\ (restricted HFD).  Our RCC-S
% calculations are of the unrestricted SCF kind.  Such difference, being
% important for open-shell systems (e.g., see comparison of different
% calculations on YbF \cite{MosyaginYbF}), is
%% , in practice,
% negligible for nondegenerate states (with closed shells) as
% is in the ground state of TlF.

\paragraph*{Acknowledgments.}
% The authors
 We
 are grateful to M.~G.~Kozlov, I.~A.~Mitropolsky, and V.~M.~Shabaev for
 valuable discussions.  This work was supported by INTAS grant No. 96--1266.
 AP, NM, TI and AT are grateful to RFBR for grants No. 99--03--33249, No.
 01--03--06334, No. 01--03--06335 and to CRDF for Grant No. 10469.  Work at TAU
 was supported by the Israel Science Foundation and the U.S.-Israel Binational
 Science Foundation.

%\samepage
%***************************************************
%\begin{widetext}
 \squeezetable
\begin{table*}[!]
\caption
 {Calculated $X$ and $M$ parameters [Eqs.\ (\ref{X}) and (\ref{M})] for the
 $^{205}$TlF ground state, compared with DHF values with different basis sets
 \protect\cite{Parpia,Quiney}.  Individual shell contributions are calculated
 from spin-averaged GRECP/SCF orbitals. GRECP/RCC-S results include spin-orbit
 interaction, and GRECP/RCC-SD values also account for electron correlation.
 All values in a.u.}
%\vspace{0.2cm}
\begin{ruledtabular}
\begin{tabular}{ll|ddd|r|dd|r}
                          & & \multicolumn{4}{c|}{$R_e=2.0844$ \AA} &
                          \multicolumn{3}{c}{$R=2.1$ \AA} \\
\hline
 \multicolumn{2}{c|}{Expansion}
 & $s,p$ & $s,p,d$ & $s,p,d,f$ & $s,p$ & $s,p$ & $s,p,d,f$ & $s,p$ \\
\hline
 \multicolumn{2}{l|}{Shell : main contribution}
 & \multicolumn{3}{c|}{$M$} & $X$ & \multicolumn{2}{c|}{$M$} & $X$    \\
\hline
 \multicolumn{2}{l|}{$1\sigma^2$ : $1s^2(F)$}
 &  0.01 &   0.02 &   0.02 &     3   &  0.00 &   0.02 &     1   \\
 \multicolumn{2}{l|}{$2\sigma^2$ : $5s^2(Tl)$}
 & -2.49 &  -2.49 & -2.49  & -1114   & -2.44 &  -2.44 & -1089   \\
 \multicolumn{2}{l|}{$3\sigma^2$ : $5p_z^2(Tl)$}
 &  4.21 &   3.91 &  3.91  &  1897   &  4.10 &   3.82 &  1851   \\
 \multicolumn{2}{l|}{$4\sigma^2$ : $2s^2(F)$}
 & -0.79 &  -0.64 & -0.64  &  -358   & -0.74 &  -0.60 &  -335   \\
 \multicolumn{2}{l|}{$5\sigma^2$ : $5d_{z^2}^2(Tl)$}
 & -0.01 &  -0.04 & -0.05  &    -2   & -0.01 &  -0.05 &    -2   \\
 \multicolumn{2}{l|}{$6\sigma^2$ : $(6s(Tl)+2p_z(F))^2$}
 & -9.38 & -10.05 & -10.06 & -4414   & -9.38 & -10.02 & -4422   \\
 \multicolumn{2}{l|}{$7\sigma^2$ : $(6s(Tl)-2p_z(F))^2$}
 & 28.13 &  27.19 &  27.19 & 12954   & 27.98 &  27.07 & 12893   \\
 \multicolumn{2}{l|}{$1\pi^4$ : $5p_x^2 5p_y^2(Tl)$}
    &  0.00 &  -0.26 &  -0.26 &     0   &  0.00 &  -0.25 &     0   \\
 \multicolumn{2}{l|}{$2\pi^4$ : $5d_{xz}^2 5d_{yz}^2(Tl)$}
    &  0.00 &   0.31 &   0.30 &     0   &  0.00 &   0.27 &     0   \\
 \multicolumn{2}{l|}{$3\pi^4$ : $2p_x^2 2p_y^2(F)$}
    &  0.00 &  -0.39 &  -0.40 &     0   &  0.00 &  -0.38 &     0   \\
 \multicolumn{2}{l|}{$1\delta^4$ : $5d_{x^2-y^2}^2 5d_{xy}^2(Tl)$}
 &  0.00 &   0.00 &  -0.02 &     0   &  0.00 &  -0.02 &     0   \\
\hline
 \multicolumn{2}{l|}{Total SCF(spin-averaged)}
             & 19.67 &  17.56 &  17.51 &  8967   & 19.52 &  17.43 &  8897   \\
\hline
 \multicolumn{2}{l|}{GRECP/RCC-S}
             & 16.12 &        &  13.84 &  9813   & 16.02 &  13.82 &  9726   \\
\hline
 DHF \cite{Parpia}&Tl:($28s28p12d8f$)  &15.61 & &     &  7743  &     &     &  \\
\hline
 DHF \cite{Quiney}&Tl:($25s25p12d8f$) &   &  &   &  &&  13.64^{\rm a}& 8098 \\
      & Tl:($28s28p14d8f$)    &     &        &   &  &&  13.62^{\rm a}& 8089 \\
      & Tl:($31s31p15d8f$)    &     &        &   &  &&  13.66^{\rm a}& 8492 \\
      & Tl:($34s34p16d9f$)    &     &        &   &  &&  13.63^{\rm a}& 8747 \\
\hline
 \multicolumn{2}{l|}{GRECP/RCC-SD}
             &       &        &  11.50 &  7635   &       &        &

\end{tabular}
\end{ruledtabular}
\vspace{0.2cm}
 \noindent $^{\rm a}  M$ is calculated in Ref.\ \cite{Quiney} using two-center
 molecular spinors, corresponding to infinite $L_{max}$ in
 Eq.~(\ref{restoration}).

\label{result}
\end{table*}
%\end{widetext}

%******************************************************

\end{document}